\documentclass[12pt, draftclsnofoot, onecolumn]{IEEEtran}
\usepackage{amssymb}
\usepackage{bbm}
\ifCLASSINFOpdf
  \usepackage[pdftex]{graphicx}
 \DeclareGraphicsExtensions{.pdf,.jpeg,.png}
\else
\fi
\usepackage[cmex10]{amsmath}
\usepackage{tikz}
\usepackage{tikz}
\usepackage{bigdelim}
\usepackage{graphicx}
\usepackage{epstopdf}
\DeclareGraphicsExtensions{.eps}

\newtheorem{thm}{Theorem}
\newtheorem{lem}{Lemma}

\newenvironment{definition}[1][Definition]{\begin{trivlist}
\item[\hskip \labelsep {\bfseries #1}]}{\end{trivlist}}

\newcommand{\qed}{\nobreak \ifvmode \relax \else
      \ifdim\lastskip<1.5em \hskip-\lastskip
      \hskip1.5em plus0em minus0.5em \fi \nobreak
      \vrule height0.75em width0.5em depth0.25em\fi}
\begin{document}
\baselineskip 8.0 mm
\title{Sum Degrees of Freedom of the $K$-user Interference Channel with Blind CSI}
\author{\begin{normalsize} Milad Johnny and Mohammad Reza Aref
\\Information System and Security Lab (ISSL),\\Sharif Universiy of Technology, Tehran, Iran\\E-mail: Johnny@ee.sharif.edu, Aref@sharif.edu
\end{normalsize}
}
\maketitle
\begin{abstract}
In this paper, we consider the problem of the interference alignment for the $K$-user SISO interference channel (IC) with blind channel state information (CSI) at transmitters. Our achievement contrary to the traditional $K-$user interference alignment (IA) scheme has more practical notions. In this case, every receiver is equipped with one reconfigurable antenna which tries to place its desired signal in a subspace which is linearly independent of interference signals. We show that if the channel values are known to the receivers only, the sum degrees-of-freedom (DoF) of the linear blind IA (BIA) with reconfigurable antenna is $\frac{Kr}{r^2-r+K}$, where $r =  \left \lceil{\frac{\sqrt{1+4K}-1}{2}} \right \rceil$. The result indicates that the optimum sum DoF for the $K-$user IC is to achieve the sum DoF of $\lim_{K \rightarrow \infty} {\frac{Kr}{r^2-r+K}}=\frac{\sqrt{K}}{2}$ for an asymptotically large interference network. Thus, the DoF of the $K$-user IC using reconfigurable antenna grows sublinearly with the number of the users, whereas it grows linearly in the case where transmitters access to the CSI. In addition, we propose both achievability and converse proof so as to show that this is the sum DoF of linear BIA with the reconfigurable antenna.
\end{abstract}


\IEEEpeerreviewmaketitle
\section{Introduction}
\IEEEPARstart{T}{he} new increasing demand for higher data rate communication motivates researchers to introduce new tools to reduce network constrains such as interference in the transmission medium. In the network area, due to the high speed of progressing, there are more opportunities for innovation and creativity to take place. Interference channel (IC) due to its important role in today's communication systems has been the focus of attention in today's wireless networks. The importance of the problem of finding the capacity of IC is so essential that after point-to-point communication scenario it is the second problem which was introduced by Shannon \cite{shannon} and it has many applications in today's communication networks. Unfortunately finding the exact capacity of the IC is so hard that it has been open for nearly half a century. While finding the exact capacity of many networks is still open, DoF or capacity pre-log can analyze capacity characteristics of such networks at high $\mathrm{SNR}$ regions. IA is a new tool and an elegant method which casts overlap shadows at the unintended receivers while the desired signals can be decoded at the intended receivers free of interference \cite{2},\cite{mad1}. Therefore, the effect of many interference signals can be reduced to a single interference signal. In \cite{2}, Cadambe and Jafar by the basic idea of IA with some constraints show that one can achieve $\frac{K}{2}$ DoF for the fast fade IC. In the perfect IA method every transmitter uses precoding matrices, which should be suitably selected to embed all the interference signals into one half of the signal space at each receiver and leave the other half without interference for the desired signal. More generally, it means that the aim of IA is to ensure that at each receiver, all the interference reaches in a signal subspace with the smallest number of dimensions and then cancels the effect of interference by zero- forcing or similar methods. Since the IA scheme provided by Cadambe and Jafar in \cite{2} is based on zero-forcing it has some degradation at low SNR regions. The performance degradation in IA networks at low SNR ratio with the assumption of CSI at transmitters was analyzed in \cite{Zhao}, in this work by the use of antenna-switching, the quality of service (QoS) at low SNR increases. Designing such precoding matrices at transmitters requires that all the transmitters have perfect access to channel state information. Unfortunately, the method of \cite{2}, for practical cases where transmitters do not have access to channel values, fails to get any achievement. The CSI was not the only barrier for implementation of such a method; the long precoder size at transmitters and the high speed of channel changing pattern show further impractical aspects of this method because such an assumption is too hard to materialize under any practical channel feedback scheme. 

Due to advantages of IA compared to trivial frequency or time division multiple access methods, there is a lot of attention to the problem of IA with imperfect CSI. Another interesting approach has developed alignment schemes that do not need instantaneous CSIT. As an example, if the channel coefficients are appropriately correlated, alignment is possible without any CSIT \cite{jafar1}, \cite{milad}. But in practical cases where channel behaviors can not be controllable, these methods fail to have a good performance. Moreover, as a forward step to study the impact of the lack of channel knowledge, \cite{milad} shows that with some constraints on the direct and interference channels, one can perfectly or imperfectly align interference; if half of the interference channel values are not available at both the transmitters and receivers, one can achieve the sum DoF of $\frac{K}{2}$. To combat the effects of imperfect CSI on IA, there are two different strategies which are related to outdated CSIT (delay CSIT) and blind CSI. 
\subsubsection{IA with delay CSIT}
In the case of delay CSIT, every transmitter has causal access to channel state information. As a first step in this regard, authors in \cite{mad2}, found the DoF rate region of MISO broadcast channel in the case of delay CSIT. Generally, they show that if a network consists of a MIMO broadcast channel with $K$ transmit antennas and $K$ receivers where each one is equipped with 1 receiver antenna, the sum DoF of $\frac{K}{1+\frac{1}{2}+\dots+\frac{1}{K}}$ is achievable. There are several works characterizing the DoF of the IC with the delayed CSIT. In \cite{8}, with the assumption of delay CSIT, it is shown that the DoF of the $K$-user IC can achieve the value of $4/(6 \, \ln(2)  - 1) \approx 1.266$ as $K\rightarrow \infty$. In this paper, the problem of IA with delay CSIT is not our objective.
\subsubsection{IA with blind CSI}
Concerning blind CSI, one basic idea to control channel coherence time and utilize partial IA is to use multi-mode switching antenna at receivers. In this case, every receiver is equipped with an antenna that can switch between different reception modes. The frame work in the case of the reconfigurable antenna is to design proper precoder and switching pattern at transmitters and receivers, respectively. The design of precoder at transmitters is independent of CSI therefore the blind IA (BIA) scheme with reconfigurable antenna only requires multi-mode antenna switching at the receivers, which does not need any significant hardware complexity \cite{Zhao} and can be easily implemented in a practical system. In \cite{Gou}, \cite{Gou1} for the MISO broadcast channel the authors show that artificially manipulating the channel itself to create the opportunities, one can facilitate BIA. They equip each user with a simple staggered antenna which can switch between multi-mode reception paths. By the use of reconfigurable antenna where the broadcast transmitter uses $M$ antennas and each receiver is equipped with multi-mode antenna switching, the network can achieve the sum DoF of $\frac{MK}{M+K-1}$ which is also the outer-bound of this channel. The authors in \cite{cwang} study the effect of zero forcing (ZF) on the method of \cite{Gou} in a cellular environment as a means for supporting downlink Multi-User MIMO (MU-MIMO) transmission. Therefore \cite{cwang}, uses similar network to MISO broadcast channel which was studied before but with specific application in the cellular environment. In \cite{Gou2}, the authors try to generalize the MISO broadcast channel of \cite{Gou} to MIMO broadcast channel with reconfigurable antenna at receivers. 

 In \cite{7}, change the network for the 3-user IC, Wang showed that in the case of blind CSI using a reconfigurable antenna at receivers the sum DoF was $\frac{6}{5}$. Our goal in this paper is to generalize the Wang's work for the case of $K-$user IC which was previously analyzed by Alaa and Ismail in \cite{Alaa}. Alaa and Ismail tried to generalize the DoF rate region of 3-user IC with the reconfigurable antenna to the $K-$user IC, but for the $K>6$ our sum DoF is larger. We show that with the aid of reconfigurable antenna at receivers, the sum DoF is $\max_{r}{\frac{Kr}{r^2-r+K}}$ where the optimum value of $r$ is a function of number of the users $K$, which is $r =  \left \lceil{\frac{\sqrt{1+4K}-1}{2}} \right \rceil$. This result indicates that when the number of the users $K$ limits to infinity, the value of $r$ goes to $\sqrt{K}$ and our BIA method can achieve sum DoF of $\frac{\sqrt{K}}{2}$ which is larger than the sum DoF upper-bound of $2$ in \cite{Alaa}, thus the sum DoF does not scale linearly with the number of users $K$ as in the case when CSI is available, but rather scales {\it sub-linearly} with the number of users.
The main contributions of this work are summarized as follows.
\begin{itemize}
\item In all parts of this paper, there is not any knowledge of CSI at the transmitters.
\item All the receivers are equipped with a simple staggered antenna switching.  This type of antenna can have several preset modes and can be performed to switch among these modes using micro-electro-mechanical switches (MEMSs) \cite{Gore}.
\item Implementing such a structure has a very low cost and is price efficient compared to original IA method. 
\item We derive an outer-bound on the sum DoF of blind IA in $K-$user IC, where each receiver uses staggered antenna switching.
\item We derive a novel achievability for the sum DoF which meets our outer-bound. 
\end{itemize}
%%%%%%%%%%%%%%%%%%%%%%%%%%%%%%%%%%%%%%%%%%%%%%%%%%%%%%%%%%%%%%%%%%%%%%%%%%%%%%%%%%%%%% 
\subsection{Organization}
This paper is organized as follows. The next section describes the system model and we present the overviews of the main result. In section III we derive a converse proof for the sum DoF of $K-$user IC. In section IV, by providing achievability, we show that our outer-bound is the sum DoF of the $K-$user IC with reconfigurable antenna at receivers. Also we provide an example for more intuition in section IV. Finally, we draw our conclusions in Section V.
%%%%%%%%%%%%%%%%%%%%%%%%%%%%%%%%%%%%%%%%%%%%%%%%%%%%%%%%%%%%%%%%%%%%%%%%%%%%%%%%%%%%%%
\subsection{Notation}
 Throughout the paper, boldface lower-case letters stand for vectors while upper-case letters show matrices. The ${\bf{A}}^\mathrm{T}$ indicates transpose operation on $\bf{A}$, the $\mathrm{tr}\{\bf{A}\}$ is defined to be sum of elements on the main diagonal of the square matrix $\bf{A}$. The $\mathrm{span}\left({\bf A}\right)$ denotes the space spanned by the columns of the matrix $\bf A$. The ${\bf{A}}_{n \times m}=\left[{\bf B},\bf{C}\right]$ means that the matrix ${\bf A}_{n \times m}$ consisted of two sub-matrices ${\bf B}_{n \times m_1}$ and ${\bf C}_{n \times m_2}$, where $m=m_1+m_2$. For the vector ${\bf v}=\left[v_1,\dots,v_n\right]^{\mathrm{T}}$, the vector ${\bf v}'=\left[{v_i,\dots,v_{i+j}}\right]^{\mathrm{T}}$ is a sub-vector of the ${\bf v}$ if $\{i,\dots,i+j\} \subseteq \{1,\dots,n\}$. Also the $\mathrm{dim}\left({\bf A}\right)$ shows the number of dimensions of the matrix $\bf{A}$. The matrix ${\bf 1}_K$ and ${\bf I}_K$ are $K \times K$ all one and identity matrices, respectively. For the square matrix ${\bf H}_{n \times n}$, ${\bf H'}={\bf H}(1:L), L\leq n$ means that $\bf H'$ is a sub-matrix of $\bf H$ where it is extracted from the first $L$ columns and the $L$ rows of the $\bf H$. The operator $\circ$ in the relation ${\bf A} \circ {\bf B}$ represents the Hadamard product between two matrices $\bf A$ and $\bf B$ with the same sizes. The $\left \lfloor.\right \rfloor$ and $\left \lceil.\right \rceil$ represent floor and ceiling operations, respectively. Also, for the set $\mathcal{C}$, $\lvert \mathcal{C} \rvert$ denotes the cardinality of the set $\mathcal{C}$. 
\section{System Model}
\begin{figure}
  \centering
  \includegraphics[width=0.5\textwidth]%
    {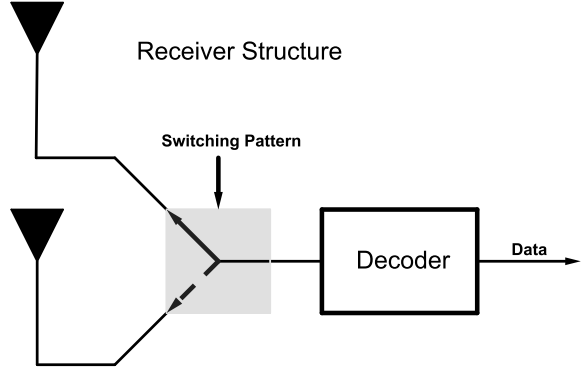}% picture filename
  \caption{Structure of the two-mode reconfigurable antenna. In this structure every receiver is equipped with two RF chains and a switch which can select between two different modes.}
\end{figure}
Consider the $K-$user IC, in this case each transmitter has one transmitter antenna. All the receivers have one reconfigurable antenna which is realized by some RF chains as shown in Figure 1 that can switch among $M$ different preset modes. Each of these RF chains (switching modes) can see a channel which is completely independent of the channel of other modes. In other words, each receiver has one antenna which can switch among $M$ different multi-mode receptions. In this case, at each time snapshot, each receiver can switch to one of the RF chains to receive its desired signal from corresponding transmitter and all other transmitters as interference signals (see Figure 1). The interference channel consists of $K$ transmitters ${\left \{\mathrm{TX}_k  \right \}_{k=1}^{K}}$ and $K$ receivers ${\left \{\mathrm{RX}_k  \right \}_{k=1}^{K}}$ which can be modeled by $ K^{2}+2K$ tuple $\left({\bf \bar{H}}^{[11]},{\bf \bar{H}}^{[12]},...,{\bf \bar{H}}^{[KK]},{\bf \bar{{x}}}^{[1]},...,{\bf \bar{{x} }}^{[K]},{\bf \bar{{y}}}^{[1]},...,{\bf \bar{{y}}}^{[K]}\right)$, where $\left({\bf \bar{{x}}}^{[1]},...,{\bf \bar{{x}}}^{[K]}\right)$
and $\left({\bf \bar{{y}}}^{[1]},...,{\bf \bar{{y}}}^{[K]}\right)$ are $K$ finite input and output of the channel respectively. In our model, the input of $\mathrm{TX}_k$ is represented by ${\bf \bar{{x}}}^{[k]}=[{x}_{1}^{[k]},....,{x}_{n}^{[k]}]^T$. Similarly the output of the channel can be represented by column matrix of ${\bf \bar{{y}}}^{[k]}=[{{y}_{1}}^{[k]},....,{{y}_{n}}^{[k]}]^T$.
The diagonal matrix ${\bf \bar{H}}^{[pq]}=\mathrm{diag}\left({\left[{h^{[pq]}_{1},h^{[pq]}_{2},\dots,h^{[pq]}_{n}}\right]}\right)$ represents channel model and maps ${\bf \bar{{x}}}^{[q]}$ to received signal at $\mathrm{RX}_p$. We can assume the received signal at the $\mathrm{RX}_p$ is consisted of $n$ time snapshot channel uses. The received signal at $\mathrm{RX}_p$ can be represented as follows:
\begin{equation}
\label{1}
{\bar{{\bf y}}^{[p]}} = \sum_{q=1}^{K} {{\bf H}^{[pq]}} {{\bf \bar{x}}^{[q]}} + {{\bf \bar{z}}^{[p]}}, \,\, p, q \in \{1,2,...,K\}
\end{equation}
where ${{\bar{\bf y}}^{[p]}}$ indicates the received signal over $n$ channel uses (time or frequency slots), ${{\bar{\bf x}}^{[q]}}$ is the transmitted signal vector by the $\mathrm{TX}_p$ subject to average power constraint of $\mathrm{SNR}$, the matrix ${\bar{ \bf{z}}}^{[p]}$ represents additive white Gaussian noise with unit power, and ${{\bar{\bf H}}^{[pq]}}$ is a diagonal matrix representing the channel coefficient between the $\mathrm{TX}_q$ and $\mathrm{RX}_p$. The channel matrix can be written as: 
\begin{equation}
\label{2}
{{\bar {\bf{H}}}^{[pq]}} = \mathrm{diag}\left( \left[ {h^{[pq]}_1,h^{[pq]}_2,\dots,h^{[pq]}_n} \right]\right),
\end{equation}
where depending on the number of antenna modes $M$, and switching pattern of RF chains at $\mathrm{RX}_p$, every $h^{[pq]}_j ,j \leq n$ can be selected from a specific set. In other words, we have:
\begin{equation}
 h^{[pq]}_j \in \lbrace{h^{[pq]}(1),h^{[pq]}(2),\dots,h^{[pq]}(M)}\rbrace.
\end{equation}
  Therefore, the channel matrix ${{\bar {\bf{H}}}^{[pq]}}$ is chosen from the set $\mathcal{H}^{[pq]}$ with the cardinality of $\lvert{\mathcal{H}^{[pq]}}\rvert=M^{n}$. In other words the diagonal matrix $\bar{\bf{H}}^{[pq]}$ as a function of ${\bf{S}}_p$ can be represented as follows:
\begin{equation}
{{\bf \bar{H}}^{[pq]}} = \mathrm{diag}([h^{[pq]}\left({S}_p {(1)}\right) \,\, h^{[pq]}\left({S}_p {(2)}\right) \,\, ... \,\, h^{[pq]}\left({S}_p {(n)}\right)]),
\end{equation}
where ${\bf{S}}_p = \left[ {S}_p {(1)}~{S}_p {(2)}~ \dots~{S}_p {(n)} \right] $ and ${S}_p {(j)} \in \{1,\dots,M\}$ shows the switching pattern matrix at $\mathrm{RX}_p$. As an example if $n=4, M=5$, ${\bf{S}}_1=\left[1,2,2,5\right]$ and the number of users $K$, we have:
\begin{equation}
{\bar{\bf H}}^{[1k]}=\mathrm{diag}\left(\left[{h^{[1k]}(1),h^{[1k]}(2),h^{[1k]}(2),h^{[1k]}(5)}\right]\right),~1 \leq k \leq K.
\end{equation}
  This switching pattern, for all channels which end in the same destination e.g. $\mathrm{RX}_p$ has the same effect. We assume that all the channel links e.g. $h^{[pq]}(S_p(j))$ between different transceivers are constant during $n$ channel uses. Therefore, the changing pattern of different channels $\bar{\bf H}^{[pq]}$ is under the control of the switching pattern of ${\bf{S}}_p$ at $\mathrm{RX}_p$. Therefore, any matrices like $\bar{\bf H}^{[pq]}$ and $\bar{\bf H}^{[pq']}$ have the same changing pattern.
  
  In all of the above relations, ${{\bf \bar{x}}^{[q]}}$ is a vector with the size of $n \times 1$ and can be represented as follows:
\begin{equation}
\label{3}
{{\bf \bar{x}}^{[q]}} = \sum_{d=1}^{d_{q}} x_{d}^{[q]} \, {{{\bf v}_{d}}^{[q]}}
\end{equation}
where $d_{q}$ is the number of symbols transmitted by the $\mathrm{TX}_q$ over $n$ channel uses, $x_{d}^{[q]}$ is the $d^{th}$ transmitted symbol and ${{{\bf v}_{d}}^{[q]}}$ is an $n \times 1$ transmit beamforming vector for the $d^{th}$ symbol. The equation of \eqref{3} can be simplified as follows:
\begin{equation}
{{\bf \bar{x}}^{[q]}}= \bar{\bf{V}}^{[q]} \bf{X}^{[q]},
\end{equation}
where ${\bf{X}}^{[q]}=\left[ x_{1}^{[q]},\dots,x_{d_q}^{[q]}\right]^{T}$ and ${\bar{\bf V}^{[q]}} = [{\bf v}_{1}^{[q]} \,\, {\bf v}_{2}^{[q]} \,\, ...\,\, {\bf v}_{d_q}^{[q]}]$. Also, ${\bar{\bf V}^{[q]}}$ is the precoder matrix at $\mathrm{TX}_q$ and ${\bf v}_{d}^{[q]}$ represents one of the basic vectors of the designed precoder at this transmitter.
\subsection{Degrees of Freedom for the $K-$user IC}
In the $K$-user IC using reconfigurable antenna at receivers with total power constraint of $\rho$, we define the degrees of freedom region as follows\cite{1}:
\begin{align}
\Bigg\{(d_1,d_2,\dots,d_K)& \in \mathbb{R}_{+}^{K}:\forall (w_1,\dots,w_K) \in \mathbb{R}_{+}^{K},& \\
&{w_1}{d_1}+\dots + {w_K}{d_K}\nonumber\leq\lim_{\rho \to \infty} \sup\left [\underset{\mathcal{\emph{R}(\rho) \in \mathcal{C}(\rho) }}{\sup}\frac {(w_1{R}_{1}(\rho)+\dots+w_K{R}_{K}(\rho))}{\log(\rho)}\right ]&\Bigg\},
\end{align}
where $\mathcal{C}(\rho) \in \mathbb{R}_{+}^{K}$ indicates the capacity region of $K-$user IC in the case of blind CSI. The sum DoF at this network can be defined by the following relation:
\begin{equation}
\mathrm{DoF}_{\mathrm{sum}}=\max{\left(\sum_{i=1}^{K}{d_i}\right)}.
\end{equation}
In the next subsection we express our main result with a theorem. In all the remaining parts of this article, we provide some tools to prove this theorem.
\subsection{Overview of the Main Result}
In this paper we explore interference alignment for the $K-$user IC with blind CSI. We provide both achievability and converse proofs on the sum DoF of the $K-$user IC with blind CSI by the aid of linear interference alignment, which, to the best of our knowledge, has not been discussed before. The summary of the results can be expressed by the following theorem.
\begin{thm}
{\it The sum DoF of the $K$-user SISO IC with BIA using reconfigurable antenna is $\max_{r}{\frac{Kr}{r^2-r+K}}, r\in \mathbb{N}$, where $r$ is a design parameter.}
\end{thm}
The term sum DoF can be maximized by setting $r=\left \lceil{\frac{\sqrt{1+4K}-1}{2}} \right \rceil$. The result indicates that when the number of users goes to infinity and there is not any information at transmitters about CSI, the value of sum DoF goes to $\frac{\sqrt{K}}{2}$. In the next section we show that by setting $r \in \mathbb{N}$ as a design parameter, the sum DoF of the  BIA in K-user IC using reconfigurable antenna is upper bounded by the term ${\frac{Kr}{r^2-r+K}}$. 

\section{Outer Bound on the Sum DoF for the BIA $K-$user IC Using staggered Antenna Switching}
In this section, we derive an upper bound on the sum DoF of the $K-$user IC with BIA using staggered antenna switching at the receivers. In all the sections of this paper, we assume no CSI at transmitters, each receiver is equipped with a reconfigurable antenna with $M$ RF chains, and each transmitter has a conventional antenna.
\subsection{Preliminary definitions and Lemma}
Before proving the converse proof of Theorem 1, we start this section by two definitions and one lemma.
\begin{definition}
The basic vector of $\bf v$ is aligned with the $\bf V$ iff ${\bf v} \in \mathrm{span}\left({\bf V}\right)$ or ${\bf v} \prec {\bf V}$.
\end{definition}
Let the basic vectors of $\mathrm{TX}_i$ be chosen from the set $\mathcal{V}^{[i]}=\{{\bf v}^{[i]}_1,\dots,{\bf v}^{[i]}_{d_i}\}$ . From Lemma 2 of \cite{7}, if two basic vectors of different transmitters e.g. ${\bf v}^{[i]}$ and ${\bf v}^{[i']}$ are aligned at a specific receiver e.g. $\mathrm{RX}_j$, since two channel matrices $\bar{\bf H}^{[ji]}$ and $\bar{\bf H}^{[ji']}$ have the same changing pattern then we should have ${\bf v}^{[i]}=\alpha {\bf v}^{[i']}$ where $\alpha$ is a scaling factor. Since the scaling factor $\alpha$ can not change the span of a vector, without losing generality one can assume $\alpha=1$. Therefore, for all the schemes regarding BIA in $K-$user interference channel using reconfigurable antenna at receiver, one should select the basic vectors of different transmitters from a common set. In other words, in the case of BIA for two transmitters e.g. $i$ and $i'$, if $\mathcal{V}^{[i]} \cap \mathcal{V}^{[i']} =\varnothing$, we can not align any basic vectors of these transmitters at any receivers. Therefore, different transmitters should choose their basic precoder vectors from the common set.   
\begin{figure}
  \centering
  \begin{tikzpicture}
  \draw (-12,4) ellipse (0.8cm and 1.5cm);
  \draw (-9,4) ellipse (0.8cm and 1.5cm);
  \draw (-12,6) circle (0.05cm);
  \node            (a) at (-12.25,6){$1$};
  \draw (-12,5) circle (0.05cm);
  \node            (b) at (-12.25,5){$l_1$};
  \draw (-12,4.5) circle (0.05cm);
  \node            (c) at (-12.25,4.5){$q_1$};
  \node            (c) at (-12.25,4){$q_2$};
  \draw (-12,4) circle (0.05cm);
  \node            (d) at (-12,3.75){$\vdots$};
  \draw (-12,3) circle (0.05cm);
  \node            (e) at (-12.25,3){$l_r$};
  \draw (-12,2) circle (0.05cm);
  \node            (f) at (-12.25,2){$q_4$};
  \draw (-12,1) circle (0.05cm);
  \node            (f) at (-12.25,1){$K$};
  \draw (-9,6) circle (0.05cm);
  \node            (g) at (-8.75,6){$1$};
  \draw (-9,5) circle (0.05cm);
  \node            (h) at (-8.75,5){$l_1$};
  \draw (-9,4.5) circle (0.05cm);
  \node            (i) at (-8.75,4.5){$q_3$};
  \draw (-9,4) circle (0.05cm);
  \node            (j) at (-9,3.75){$\vdots$};
  \draw (-9,3) circle (0.05cm);
  \node            (k) at (-8.75,3){$l_r$};
  \draw (-9,2) circle (0.05cm);
  \node            (l) at (-8.75,2){$q_4$};
  \draw (-9,1) circle (0.05cm);
  \node            (l) at (-8.75,1){$K$};
  \node            (m) at (-7,4){RX set ${\mathcal{L}}^{t}$};
  \node            (n) at (-14,4){TX set ${\mathcal{L}}^{t}$};
  \node            (o) at (-12.25,0){\begin{LARGE}
  TX
  \end{LARGE}};
  \node            (p) at (-9,0){\begin{LARGE}
  RX
  \end{LARGE}};
  \node            (tk) at (-11.75,1){};
  \node            (rk) at (-9.25,1){};
  \node            (t1) at (-11.75,2){};
  \node            (tlq) at (-11.75,4){};
  \node            (tq4) at (-11.75,6){};
  \node            (r1) at (-9.25,2){};
  \node            (rlq) at (-9.25,4){};
  \node            (rq4) at (-9.25,6){};
  \draw[->] (t1) edge (r1);
  \draw[->] (t1) edge (rlq);
  \draw[->] (t1) edge (rq4);
  \draw[->] (tlq) edge (r1);
  \draw[->] (tlq) edge (rlq);
  \draw[->] (tlq) edge (rq4);
  \draw[->] (tq4) edge (r1);
  \draw[->] (tq4) edge (rlq);
  \draw[->] (tq4) edge (rq4);
  \draw[->] (t1) edge (rk);
  \draw[->] (tk) edge (r1);
  \draw[->] (tk) edge (rk);    
  \end{tikzpicture}
  \caption{In this figure we show transceivers number of the set ${\mathcal{L}}^{t}=\{l_1,l_2,\dots,l_r\}$ with the closed circular shape. The complimentary transceivers out of this circular shape can be modeled by the set $\{1,\dots,K\}-{\mathcal{L}}^{t}$. Also there is a connection among all transmitters and receivers, but to avoid being so crowded we show a few of them \cite{milad}.}
\end{figure}
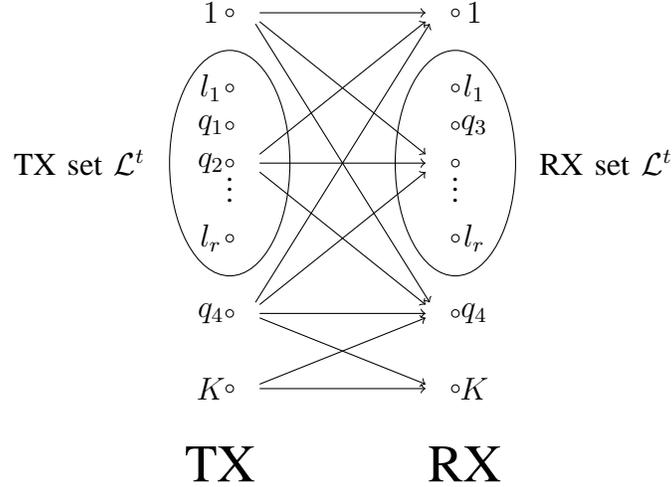
 
\textit{{Remark:}} At $\mathrm{RX}_q$ the basic vector of $\bar{{\bf{H}}}^{[qq]}{\bf{v}}^{[q]}$ should not be in the span of the $\bar{{\bf{H}}}^{[qq']}\bar{{\bf{V}}}^{[q']}$ otherwise, the desired signal space is polluted by interference of $\mathrm{TX}_{q'}$.

 Now consider the set ${\mathcal{L}}^{t}=\{l_1,\dots,l_r\}\subseteq \{1,\dots,K\}$ where $\vert {\mathcal{L}}^{t} \vert = r$ and $1 \leq t \leq \binom{K}{r}$. The set ${\mathcal{L}}^{t}$ shows the index of the subset of the transmitters. The following lemma limits the selection of joint vectors between different transmitters and is the starting point of our converse proof.
\begin{lem}
If ${\bf{v}}^{[q]}$ is aligned with the interference of the $r-1$ transmitters $\mathrm{TX}_j$, $j \in \mathcal{L}^{t}-\{q\}$ at $\mathrm{RX}_{j'}$, $j' \in \{1,\dots,K\}-\mathcal{L}^{t}$, it can not be aligned with the interference generated by the $\mathrm{TX}_j$ at $r-1$ receivers of the set $\mathcal{L}^{t}-\{q\}$.
\end{lem}
\begin{IEEEproof}
For a better intuition see Figure 2 and suppose that $\mathrm{TX}_{q_1}$ and $\mathrm{TX}_{q_2}$ are two arbitrary transmitters  where, $q_1,q_2 \in \mathcal{L}_t$. Also the $\mathrm{RX}_{q_3}, q_3 \in \mathcal{L}^t$ and the $\mathrm{RX}_{q_4}, q_4 \in \{1,\dots,K\}-\mathcal{L}^t$ are two arbitrary receivers. From the assumption of this lemma we can assume:
\begin{equation}
\bar{{\bf{H}}}^{[q_4q_1]}{\bf{v}}^{[q_1]} \in \text{span}\left({\bar{{\bf{H}}}^{[q_4 q_3]}\bar{{\bf{V}}}^{[q_3]}}\right).
\end{equation}
From Lemma 2 of \cite{7}, since $\bar{{\bf{H}}}^{[q_4 q_1]}$ and $\bar{{\bf{H}}}^{[q_4 q_3]}$ are diagonal and have the same changing pattern, ${\bf{v}}^{[q_1]} \in \text{span}\left({\bar{{\bf{V}}}^{[q_3]}}\right)$.\\
(Proof by contradiction.) We take the negation of our lemma and suppose it is true. Assume, to the contrary, that:
\begin{equation}
\left\{ \exists q_3 \in \mathcal{L}^t: \text{span}\left( \bar{{\bf{H}}}^{[q_{3}q_{1}]}{\bf{v}}^{[q_1]}\right)\in \text{span}\left( \bar{{\bf{H}}}^{[q_{3}q_{2}]}{\bar{{\bf{V}}}}^{[q_2]}\right) \right\}.
\end{equation}   
Then, we have:
\begin{equation}
\text{span}\left({\bar{{\bf{H}}}}^{[q_3 q_1]}{\bf{v}}^{[q_1]}\right) \in \text{span} \left({\bar{{\bf{H}}}^{[q_3 q_3]}} \left(\bar{{\bf{H}}}^{[q_3 q_3]}\right)^{-1}\bar{{\bf{H}}}^{[q_3 q_2]}{\bar{{\bf{V}}}^{[q_2]}}\right).
\end{equation}
Since ${\bar{{\bf{H}}}^{[q_3 q_1]}}$ and ${\bar{{\bf{H}}}^{[q_3 q_3]}}$ have the similar changing pattern, we get:
\begin{equation}
\text{span}\left({\bf{v}}^{[q_1]}\right) \in \text{span} \left(\left(\bar{{\bf{H}}}^{[q_3 q_3]}\right)^{-1}\bar{{\bf{H}}}^{[q_3 q_2]}{\bar{{\bf{V}}}^{[q_2]}}\right).
\end{equation}
Therefore, since ${\bf{v}}^{[q_1]} \in \text{span}\left({\bar{{\bf{V}}}^{[q_3]}}\right)$, we have:
\begin{equation}
\text{dim} \left( {\bar{{\bf{V}}}^{[q_{3}]}} \cap \left(\bar{{\bf{H}}}^{[q_{3} q_{3}]}\right)^{-1}\bar{{\bf{H}}}^{[q_{3} q_{2}]}{\bar{{\bf{V}}}^{[q_{2}]}} \right) >0,
\end{equation}
and finally we get:
\begin{equation}
\text{dim} \left( \bar{{\bf{H}}}^{[q_{3} q_{3}]}{\bar{{\bf{V}}}^{[q_{3}]}} \cap \bar{{\bf{H}}}^{[q_{3} q_{2}]}{\bar{{\bf{V}}}^{[q_{2}]}} \right) >0.
\end{equation}
The above relation shows that the desired signal $\bar{{\bf{H}}}^{[q_{3} q_{3}]}{\bar{{\bf{V}}}^{[q_{3}]}}$ at $\mathrm{RX}_{q_{3}}$ has been polluted by the interference of $\mathrm{TX}_{q_{2}}$. Hence by the assumption of $\left\{ \exists q_3 \in \mathcal{L}^t: \text{span}\left( \bar{{\bf{H}}}^{[q_{3}q_{1}]}{\bf{v}}^{[q_1]}\right)\in \text{span}\left( \bar{{\bf{H}}}^{[q_{3}q_{2}]}{\bar{{\bf{V}}}}^{[q_2]}\right)\right\}$ we have a contradiction. This contradiction shows that the given assumption is false and the statement of the lemma is true. So, this completes the proof.
\end{IEEEproof}

Therefore, every basic vector of each transmitter aligns with interference generated from $r-1$ transmitters at $K-r$ receivers. In other words, if ${\bf{v}}^{[q]}$ is one of the basic vectors of $\mathrm{TX}_q$, we have:
\begin{equation}
\bar{{\bf{H}}}^{[pq]}{\bf{v}}^{[q]} \prec \bar{\bf{H}}^{[pq']}\bar{{\bf{V}}}^{[q']},
\end{equation}
where, $(q, q' \in \mathcal{L}^t,~q \neq q'$ and $p \in \{1,\dots,K\}-\mathcal{L}^t$.
\begin{definition}
$d_{{i_1}{i_2}\dots{i_r}},~i_1 \neq i_2 \neq \dots \neq i_r$ shows the number of dimensions which is occupied by transmitters ${\mathrm{TX}}_{i_1}$, ${\mathrm{TX}}_{i_2}$,... and ${\mathrm{TX}}_{i_r}$ at $\mathrm{RX}_j$, where $j\notin \{i_1,i_2,\dots,i_r\}$. In other words, $d_{{i_1}{i_2}\dots{i_r}}=\lvert {\cap _{i_1,\dots,i_r} {\mathcal{V}^{[i]}}} \rvert ,~i_1 \neq i_2 \neq \dots \neq i_r$. 
\end{definition}
From above definition it is straightforward to show that for every permutation of $i'_1,\dots,i'_r \in \{i_1,i_2,\dots,i_r\}$ we have:
\begin{equation}
d_{{i_1}{i_2}\dots{i_r}}=d_{{i'_1}{i'_2}\dots{i'_r}}.
\end{equation} 
\subsection{Converse Proof:}
The converse proof follows from the following upper bound on the DoF of the $K-$user interference channel with BIA.
At $\mathrm{RX}_j$ receiver the interference signal from transmitters $\mathrm{TX}_{i_1}$,$\mathrm{TX}_{i_2}$,... and $\mathrm{TX}_{i_r}$, where $j\notin \{i_1,i_2,\dots,i_r\}$ jointly occupy $d_{{i_1}{i_2}\dots{i_r}}$ dimensions. In other words, every shared vectors between $r$ different transmitters ($\mathrm{TX}_{i_1}$,$\mathrm{TX}_{i_2}$,... and $\mathrm{TX}_{i_r}$) occupy just only one dimension at $\mathrm{RX}_j$. On the other hand the total number of dimensions at each receiver is $n$. Therefore, at the $\mathrm{RX}_j$ we have:
\begin{equation}
\label{14}
d_1+d_2+\dots+d_K-\left(r-1\right) \sum_{i_1} \dots \sum_{i_r}{d_{i_1,\dots,i_r}} \leq n,~i_1,\dots,i_r \in \{1,\dots,K\}-\{j\},
\end{equation}
where, the coefficient $(r-1)$ comes from this fact that $d_{i_1,\dots,i_r},~i_1,\dots,i_r \in \{1,\dots,K\}-\{j\}$ just only occupies one dimension at $j^{th}$ receiver while it counts $r$ times in the term $d_1+d_2+\dots+d_K$. Similarly at all the receivers we have:
\begin{equation}
\label{15}
\begin{aligned}
&\mathrm{RX}_{1}:~d_1+d_2+\dots+d_K-\left(r-1\right) \sum_{i_1} \dots \sum_{i_r} {d_{i_1,\dots,i_r}} \leq n,~i_1,\dots,i_r \in \{1,\dots,K\}-\{1\}&\\
&\mathrm{RX}_{2}:~d_2+d_1+\dots+d_K-\left(r-1\right) \sum_{i_1} \dots \sum_{i_r}{d_{i_1,\dots,i_r}} \leq n,~i_1,\dots,i_r \in \{1,\dots,K\}-\{2\}&\\
&\vdots&\\
&\mathrm{RX}_{K}:~d_K+d_1+\dots+d_{K-1}-\left(r-1\right)\sum_{i_1} \dots \sum_{i_r} {d_{i_1,\dots,i_r}} \leq n,~i_1,\dots,i_r \in \{1,\dots,K\}-\{K\}.&\\
\end{aligned}
\end{equation}
Adding all the above relations we conclude that:
\begin{equation}
\label{DOFlower}
K \sum_{i=1}^{K}{d_i}-\left(K-r\right)\left(r-1\right)\sum_{i_1=1}^{K} \dots \sum_{i_r=1}^{K} {d_{i_1,\dots,i_r}} \leq Kn,~i_1,\dots,i_r \in \{1,\dots,K\}
\end{equation} 
The term $(K-r)$ comes from this fact that $\sum_{i_1=1}^{K} \dots \sum_{i_r=1}^{K}{d_{i_1,\dots,i_r}}, i_j \neq k $ consists of $\binom{K-1}{r}$ summation while $\sum_{i_1=1}^{K} \dots \sum_{i_r=1}^{K}{d_{i_1,\dots,i_r}}$ consists of $\binom{K}{r}$ summation. Therefore, the term $\frac{K \binom{K-1}{r}}{\binom{K}{r}}=\left(K-r\right)$ comes in to our inequality of \eqref{DOFlower}. In addition, since every shared dimension e.g. $d_{i_1,\dots,i_r}$ has been shared between $r$ different transmitters we have:
\begin{equation} 
r \sum_{i_1=1}^{K} \dots \sum_{i_r=1}^{K}{d_{i_1,\dots,i_r}} \leq \sum_{i=1}^{K}{d_i}.
\end{equation}
Therefore from \eqref{DOFlower} we have:
\begin{equation}
\label{DOFlower2}
K \sum_{i=1}^{K}{d_i}-\frac{\left(K-r\right)\left(r-1\right)}{ r } \sum_{i=1}^{K}{d_i} \leq Kn.
\end{equation}
After simplifying \eqref{DOFlower2} we get:
\begin{equation}
\frac{\sum_{i=1}^{K}{d_i}}{n} \leq \frac{Kr}{r^2-r+K},
\end{equation}
thus, the converse proof completed.\\
In order to find the maximum value of the upper-bound on the sum DoF, we analyze the continuous function of $f(x)=\frac{Kx}{x^2-x+K}$. The first derivation of this function has just one positive root of $x=\sqrt{K}$ which shows that it has just only one extremum point. Also it can easily be shown that for $x\geq 0$ the function $f(x)$ is greater than or equal to zero. Since $f(x=0)=0$ and $f(x\rightarrow \infty ) \rightarrow 0^{+}$.
%\begin{figure}
%  \centering
%  \includegraphics[width=1\textwidth]%
%    {fig2.eps}% picture filename
%  \caption{The function $f(x)=\frac{Kx}{x^2-x+K}$ versus continuous variable of $x$ for $K=4$.}
%\end{figure} 
Therefore, the maximum value of the $d(r)$ can be calculated by finding out the minimum value of $r\in \mathbb{N}$ such that:
\begin{equation}
d(r+1)-d(r)\leq 0.
\end{equation} 
In order to find $r$ to satisfy $d(r+1)-d(r)\leq 0$ condition we have:
\begin{align}
d(r+1)-d(r)&=\frac{K(r+1)}{\underbrace{(r+1)^2-(r+1)+K}_{>0}}-\frac{Kr}{\underbrace{r^2-r+K}_{>0}}&\\
&=\frac{K(r+1)(r^2-r+K)-Kr\left((r+1)^2-(r+1)+K \right)}{\underbrace{\left((r+1)^2-(r+1)+K\right)\left(r^2-r+K\right)}_{>0}}\\
&=\frac{-K\left( r^2+r-K \right)}{\underbrace{\left((r+1)^2-(r+1)+K\right)\left(r^2-r+K\right)}_{>0}} \leq 0&\\
&\Rightarrow r \geq \frac{\sqrt{1+4K}-1}{2},&
\end{align}
Therefore, the minimum value of $r \in \mathbb{N}$ which satisfies above equation is $ r^{*}=\left \lceil{\frac{\sqrt{1+4K}-1}{2}} \right \rceil$.
Thus, for a large number of users, the sum DoF of BIA in the $K$-user interference channel approaches $\frac{\sqrt{K}}{2}$. In the following section, we propose an algorithm to systematically generate the antenna switching patterns and the beamforming vectors such that the $\frac{Kr}{r^2-r+K}$ sum DoF is achieved.

\section{Achievable DoF Using Staggered Antenna Switching}
In the previous section we derived an upper-bound on the sum DoF of the $K-$user IC with blind CSI. As we discussed in the system model, the transmitters and the receivers should design proper beamforming vectors and switching patterns, respectively to align maximum dimension of the interference signals at their receivers. From the previous section we found out all the transmitters should use some shared basic vectors at their transmitters. These basic vectors for implementation should satisfied following constrains:  
\begin{itemize}
\item {{\it{Constrain 1:}} The shared basic vector ${\bf v}^{[p]}_i= \bigcap_{p \in \{p_1,\dots,p_r\}}{{\mathcal{V}}^{[p]}}$ which is used commonly at $\{{\mathrm{TX}}_{p_1},\dots,{\mathrm{TX}}_{p_r}\}$ after being multiplied by ${\bf{\bar{H}}}^{[lm]}, l \in \{1,\dots,K\}-\{p_1,\dots,p_r\}, m \in \{ p_1,\dots,p_r\}$ should be aligned at their complimentary receivers $\mathrm{RX}_l, l \in \{1,\dots,K\}-\{p_1,\dots,p_r\}$.}
\item {{\it{Constrain 2:}} The shared basic vector ${\bf v}^{[p]}_i= \bigcap_{p \in \{p_1,\dots,p_r\}}{{\mathcal{V}}^{[p]}}$ which is used commonly at $\{{\mathrm{TX}}_{p_1},\dots,{\mathrm{TX}}_{p_r}\}$ after being multiplied by ${\bf{\bar {H}}}^{[lm]}, l, m \in \{ p_1,\dots,p_r\}$ channel matrices should be linearly independent of each other at their corresponding receivers $\mathrm{RX}_l, l \in \{p_1,\dots,p_r\}$.}
\end{itemize}
The first constrain is the reduce the effect of interference signals at interference paths (IA at interference paths) and the second constrain is to separability of desired signal or the condition that the desired signal space can be subtracted from interference signals space (desired signal decodability). 
\begin{definition}
Assume that all the rows of the matrix ${\bf A}_{L \times K}$ are selected from the set of $\mathcal{A}=\{{{\bf a'}_1},\dots,{{\bf a'}_N}\}$, we say that matrix ${\bf A}_{L \times K}=\left[{{\bf a}_1}^{\mathrm{T}},\dots,{{\bf a}_L}^{\mathrm{T}}\right]^{\mathrm{T}}$ has the maximum distinct rows on the set of $\mathcal{A}$ if $\lvert{\{{{\bf a}_1},\dots,{{\bf a}_N}\} \cap \mathcal{A}}\rvert$ is maximized.
\end{definition}
We design both the precoder matrices and switching patterns from the basic matrix of $\bf{{F}}$. In other words, based on matrix $\bf{F}$ one can design proper precoders and switching patterns at transmitters and receivers respectively. The basic matrix ${\bf{{F}}} \in {\{0,1\}}^{n\times K}$ has the following form:
\begin{equation}
\label{Frelation}
{\bf {F}}^{\mathrm{T}} = \left[{\underbrace{{\bf A},\dots,{\bf A}}_{\text{r-1 times}}, {{\bf B}^{\mathrm{T}}_{(n-(r-1)K)\times K}}}\right] 
\end{equation}
where, $n=\binom{K-1}{r}+r\binom{K-1}{r-1}$, ${\bf A}={\bf 1}_{K \times K}-{\bf I}_{K \times K}$ and ${\bf B}_{(n-(r-1)K)\times K}$ is a matrix with $(n-(r-1)K)$ rows. Consider $\mathcal{B}$ is a set with $\lvert \mathcal{B} \rvert = \binom{K}{K-r}$, each member of the set $\mathcal{B}$ is a vector with the length of $K$ and its members contain exactly $K-r$ ones and $r$ zeros. Referring to the definitions, the matrix ${\bf B}_{(n-(r-1)K)\times K}$ has the maximum distinct rows on the set $\mathcal{B}$. Also ${\bf 1}_{K \times K}$ is an all-ones square matrix and ${\bf I}_{K \times K}$ is an identity matrix. For instance, in the case of $K = 4$ and $r=3$, the matrix ${\bf F}$ can be represented as follows (take note $r=3$ is not the optimum value for the $K=4$):
\begin{equation}
\label{fexam}
{\bf {F}}^{\mathrm{T}}=
\left[ \begin{array}{cccccccccccc}
  0&1&1&1&0&1&1&1&1&0&0&0\\
  1&0&1&1&1&0&1&1&0&1&0&0\\
  1&1&0&1&1&1&0&1&0&0&1&0\\
  1&1&1&0&1&1&1&0&0&0&0&1
\end{array} \right].
\end{equation}
 The matrix $\bf F$ consists of $K$ columns where $j^{th}$ column of this matrix is expressed by $\bf {F}_j$. We continue this section by designing beamforming vectors at transmitters.
\subsection{Beamforming vectors generation}
 To design beamforming vectors, we assume all the elements of the beamforming vectors are binary, thus ${\bf v}_{d}^{[i]}(j) \in \{0,1\}$. In this case all the basic column vectors of the precoder matrix $\bar{\bf{V}}^{[p]}$ at $\mathrm{TX}_{p}$ are chosen from the following set:
\begin{equation}
\label{preco}
\mathcal{V}^{[p]}=\Big{\lbrace} {{{\bf {F}}_{i_1}} \circ {{\bf {F}}_{i_2}} \circ ... \circ {{\bf {F}}_{i_{K-r}}} \Big{\vert} ~i_l \in \{1,\dots,K\}-\{p\}} \Big{\rbrace}.
\end{equation}
It means that $\lvert \mathcal{V}^{[p]} \rvert = \binom{K-1}{K-r}$ and therefore all the precoder matrices have the size of $n \times \binom{K-1}{K-r}$ or equivalently have the size of $n \times  \binom{K-1}{r-1}$.
Thus every $r$ different transmitter e.g. $\mathrm{TX}_{q_1},~\mathrm{TX}_{q_2},\dots$ and $\mathrm{TX}_{q_r}$ has exactly one shared basic vector. In other words we have:
\begin{equation}
\label{precoderdesign}
\Big{\vert} {\bigcap_{q_1,\dots,q_r}{\mathcal{V}^{[q]}}} \Big{\vert} = 1.
\end{equation}
Also from \eqref{preco} and \eqref{precoderdesign}, we can conclude that every shared basic vector among the transmitters of the set $\mathcal{Q}=\{q_1,\dots,q_r\}$ can be represented as follows:
\begin{equation}
\label{preshare}
\Big{\vert} {\bigcap_{q \in \mathcal{Q}}{\mathcal{V}^{[q]}}} \Big{\vert}=\Big{\lbrace} {{{\bf {F}}_{q'_1}} \circ {{\bf {F}}_{q'_2}} \circ ... \circ {{\bf {F}}_{q'_{K-r}}} \Big{\vert} ~q'_l \in \{1,\dots,K\}-\mathcal{Q}} \Big{\rbrace}.
\end{equation}
 In the next subsection we discuss how to design proper encoders at each receiver.
\subsection{Antenna Switching Pattern at the Receivers}
As it was declared in section II, each receiver is equipped with a multi-mode antenna which can select among $M$ different receiving paths. Therefore, for the switching pattern ${\bf {S}}_p={\left[{S_p (1),\dots,S_p (n)}\right]}^{\mathrm{T}}$ where $S_p (j) \in \{0,\dots,M-1\}$ we should find proper ${\bf {S}}_p$ among $M^{n}$ different switching patterns to satisfy alignment constraints.
Therefore we define a switching matrix $\bf {S}$ which is an $n \times K$. Based on this matrix all the switching patterns at different receivers are designed. We define the matrix $\bf {S}$ as follows:
\begin{equation}
\label{switchingpattern}
{\bf {S}}^{\mathrm{T}} = \left[{{\bf {A}},{\bf {A}}+2{\bf I}_{K \times K},\dots,{\bf {A}}+(r-1){\bf I}_{K \times K}, {\bf B}^{\mathrm{T}}_{(n-(r-1)K)\times K}}\right]. 
\end{equation}
Now, let ${\bf{S}}_p$ be the antenna switching pattern at $\mathrm{RX}_p$. This switching pattern is equal to $p^{th}$ column of the matrix $\bf{S}$. In other words if the matrix $\bf S$ is represented as:
 \begin{equation}
{\bf {S}}=
\left[ \begin{array}{cccc}
  s_{11}&s_{12}&\dots&s_{1K}\\
  s_{21}&s_{22}&\dots&s_{2K}\\
  \vdots&\vdots&\dots&\vdots\\
  s_{n1}&s_{n2}&\dots&s_{nK}
\end{array} \right],
\end{equation}
the switching pattern at $\mathrm{RX}_p$ can be calculated as follows:
\begin{equation}
{\bf{S}}_p = {\left[ s_{1p}, s_{2p}, \dots ,s_{np} \right]}^{\mathrm{T}},
\end{equation}
where $s_{ip}$ indicates $i^{th}$ row and $p^{th}$ column of the matrix $\bf S$. As it is clear from \eqref{switchingpattern}, all the elements of the matrix $\bf S$ are in the set of $\mathcal{P}=\{0,\dots,r-1\}$. It shows that in our switching pattern design we use an antenna with $\lvert \mathcal{P} \rvert=r$ different reconfigurable modes. Therefore, in the designed switching pattern each receiver has been equipped with single antenna with $M=r$ different receiving modes.
\subsection{Analyzing designed precoders at transmitters}
Now we must show that all the basic vectors generated at a specific transmitter e.g. $\mathrm{TX}_q$ are linearly independent. As it is shown in \eqref{Frelation}, the matrix $\bf F$ has a repetitive structure and since all the basic vectors of the different transmitters generated from Hadamard product of different columns of the matrix $\bf F$, they also have the same structure of $\bf F$. Therefore, every basic vector like ${\bf{v}}^{[q]}_{i} \in \mathcal{V}^{[q]},1 \leq i \leq \binom{K-1}{r-1}$ can be equivalently expressed by $r$ sub-matrices as follows:
\begin{equation}
\label{vectorsubmatrix}
{\bf{v}}^{[q]}_{i}={\left[{\left({\bf{v}}^{[q]}_{{{\bf e_1} i}}\right)^{\mathrm{T}}},\dots,{\left({\bf{v}}^{[q]}_{{{\bf e_{r-1}} i}}\right)^{\mathrm{T}}},{\left({\bf{v}}^{[q]}_{{{\bf f} i}}\right)^{\mathrm{T}}}\right]}^{\mathrm{T}},
\end{equation}
where all the vectors of the set $\{{\bf{v}}^{[q]}_{{{\bf e_1} i}},\dots,{\bf{v}}^{[q]}_{{{\bf e_{r-1}} i}}\}$ are similar and have the same size of $K \times 1$. Since the basic vector of ${\bf{v}}^{[q]}_{{{\bf e_1} i}}=\left[v^{[q]}_1,\dots,v^{[q]}_K\right]^{\mathrm{T}},$ generated from Hadamard product of $K-r$ column of the matrix ${\bf A}^{\mathrm{T}}={\left({\bf 1}_{K \times K}-{\bf I}_{K \times K}\right)}^{\mathrm{T}}$, we can conclude that exactly $K-r$ elements of the vector ${\bf{v}}^{[q]}_{{{\bf e_1} i}}$ are zero and $r$ elements of this vector are ones. Similarly, basic vector of ${\bf{v}}^{[q]}_{{{\bf f} i}}$ with the size of $\left({n-(r-1)K}\right) \times 1$ generated from Hadamard product of the different combination of the columns of the matrix $\bf B$. Similar notion can be expressed for the basic vector of ${\bf{v}}^{[q]}_{{{\bf f} i}}=\left[v^{[q]}_1,\dots,v^{[q]}_{\left({n-(r-1)K}\right)}\right]^{\mathrm{T}}$ but with a different result. The following lemma shows that all generated basic vectors from \eqref{precoderdesign} at a specific transmitter are linearly independent. 
\begin{lem}
{\it For all the values of the $K$ and $r=\lceil{\frac{\sqrt{1+4K}-1}{2}}\rceil$, all the basic vectors with the designed algorithm at a specific transmitter e.g. $\mathrm{TX}_p$ are linearly independent.}
\end{lem}
\begin{IEEEproof}
Consider $\mathrm{TX}_p$, all the basic vectors of this transmitter are chosen from the following set:
\begin{equation}
\mathcal{V}^{[p]}=\Big{\lbrace} {{{\bf {F}}_{i_1}} \circ {{\bf {F}}_{i_2}} \circ ... \circ {{\bf {F}}_{i_{K-r}}} \Big{\vert} ~i_l \in \{1,\dots,K\}-\{p\}} \Big{\rbrace}.
\end{equation} 
We must show that at $\mathrm{TX}_{p}$ where $\bar{{\bf{V}}}^{[p]}=\left[{{\bf {v}}^{[p]}_{1},\dots,{\bf {v}}^{[p]}_{\binom{K-1}{r-1}}}\right] $, all the vectors of the ${\bf {v}}^{[p]}_{1},\dots$ and ${\bf {v}}^{[p]}_{\binom{K-1}{r-1}}$ are linearly independent. Since ${\bf{v}}^{[p]}_{{{\bf f} i}}, 1 \leq i \leq \binom{K-1}{r-1}$ is a sub-vector of all precoder vectors, if we show that all these basic vectors are linearly independent we can conclude that all the basic vectors of the set $\mathcal{V}^{[p]}$ are linearly independent. The basic vectors of ${\bf{v}}^{[p]}_{{{\bf f} i}}, 1 \leq i \leq \binom{K-1}{r-1}$ are generated from Hadamard product of the columns of the matrix ${\bf{B}}_{(n-(r-1)K)\times K}$. As it is defined in \eqref{Frelation} each row of the matrix ${\bf{B}}_{(n-(r-1)K)\times K}$ contains exactly $K-r$ ones. 

It is completely straight forward to show that for every value of $K$ and $r=\lceil{\frac{\sqrt{1+4K}-1}{2}}\rceil$ the value of $\left(n-(r-1)K\right)-\binom{K-1}{r-1} \geq 0$. Since each row of the matrix $\bf B$ has exactly $K-r$ ones, all the basic vectors ${\bf{v}}^{[p]}_{{\bf f}_ {i_l}}, 1 \leq i_l \leq \binom{K-1}{r-1}$ referring to equation \eqref{Frelation} and \eqref{preco}, at least have a nonzero element in the unique position.  Therefore, for all values of $K$ and $r=\lceil{\frac{\sqrt{1+4K}-1}{2}}\rceil$ all the generated ${\bf{v}}^{[p]}_{{\bf f}_ {i_l}}, 1 \leq i_l \leq \binom{K-1}{r-1}$ are linearly independent. Since ${\bf{v}}^{[p]}_{{\bf f}_ {i_l}}, 1 \leq i_l \leq \binom{K-1}{r-1}$ are the sub-vectors of the basic vectors of ${\bf v}^{[p]}_{i_l}, 1 \leq i_l \leq \binom{K-1}{r-1}$, all these basic vectors are linearly independent too. Therefore, the proof was completed.  
%\IEEEQEDhere
\end{IEEEproof}
\begin{lem}
{\it For the basic vectors of ${\bf v}^{[q]}_i=\bigcap _{q \in \mathcal{Q}}{\mathcal{V}^{[q]}}, \mathcal{Q}=\{q_1,\dots,q_r\}$, using switching pattern ${\bf S}_p$ at $\mathrm{RX}_p, p \notin \mathcal{Q}$, the received basic vectors of $\bar{{\bf H}}^{[pq]}{\bf v}^{[q]}_i, q \in \mathcal{Q}$ at $\mathrm{RX}_p$ are aligned with each other.}
\end{lem}
\begin{IEEEproof}
The proof was provided by analyzing both nonzero elements of the basic vector ${\bf v}^{[q]}_i$ and the structure of the diagonal matrix $\bar{\bf H}^{[pq]}$. Similar to \eqref{vectorsubmatrix}, the basic vector of ${\bf v}^{[q]}_i, q \in {\mathcal{Q}}$ can be represented by the $r$ sub-matrices as follows:
\begin{equation}
\label{vectorrep}
{\bf{v}}^{[q]}_{i}={\left[{\left({\bf{v}}^{[q]}_{{{\bf e_1} i}}\right)^{\mathrm{T}}},\dots,{\left({\bf{v}}^{[q]}_{{{\bf e_{r-1}} i}}\right)^{\mathrm{T}}},{\left({\bf{v}}^{[q]}_{{{\bf f} i}}\right)^{\mathrm{T}}}\right]}^{\mathrm{T}}.
\end{equation} 
From \eqref{preshare} and the structure of matrix ${\bf A}$, for the sub-vector of ${\bf{v}}^{[q]}_{{{\bf e} i}}={\left[{{v^{[q]}_{{{\bf e} i}}}(1),{v^{[q]}_{{{\bf e} i}}}(2),\dots,{v^{[q]}_{{{\bf e} i}}}(K)}\right]}^{\mathrm{T}}$, $ q \in {\mathcal{Q}}$ we have:
\begin{equation}
{v^{[q]}_{{{\bf e} i}}}({q_1})={v^{[q]}_{{{\bf e} i}}}({q_2})=\dots={v^{[q]}_{{{\bf e} i}}}({q_r})=1.
\end{equation}
It means that the only nonzero elements of ${\bf{v}}^{[q]}_{{{\bf e} i}}$ are its $\{ {q_1}^{th},{q_2}^{th},\dots,{q_r}^{th} \}$ elements where the switching pattern ${\bf S}_p$ at $\mathrm{RX}_p$ has the value of one. Similarly for the nonzero elements ${\bf{v}}^{[q]}_{{{\bf f} i}}=\left[{{v}^{[q]}_{{\bf f} i}(1)},\dots,{{v}^{[q]}_{{\bf f} i}(n-(r-1)K)}\right]^{\mathrm{T}}$ e.g. ${{v}^{[q]}_{{\bf f} i}(j)}=1$ the value of ${\bf S}_p(j+(r-1)K)$ is equal to 1. Therefore, at $\mathrm{RX}_p, p \in \{1,\dots,K\}-\mathcal{Q}$ all the basic vectors like ${\bf{v}}^{[p]}_{i}, p \in \{p_1,\dots,p_r\}$ received by multiplying the constant number of $h^{[qp]}\left(1\right)$ at $\mathrm{RX}_q$. Thus all the ${\bar{\bf{H}}}^{[pq]}{\bf v}^{[q]}_i$, $q \in \mathcal{Q}$ and $p \in \{1,\dots,K\}-\mathcal{Q}$ arrive along  the basic vector of ${\bf v}^{[q]}_i$. So the proof is completed.
\end{IEEEproof}
 As an example for Lemma 3 and better intuition, consider the structure of $\bf F$ in relation \eqref{fexam}, the following analysis can be applied for different shared basic vectors.
\begin{itemize}
\item The shared basic vector among $\mathrm{TX}_1$, $\mathrm{TX}_2$ and $\mathrm{TX}_3$ can be represented as follows ($\mathcal{Q}=\{1,2,3\}$):
\begin{equation}
\Big{\vert} {\bigcap_{q \in \mathcal{Q}}{\mathcal{V}^{[q]}}} \Big{\vert}=\Big{\lbrace} {{{\bf {F}}_{q'_1}} \Big{\vert} ~q'_l \in \{1,\dots,4\}-\mathcal{Q}} \Big{\rbrace}=\{{{\bf {F}}_{4}}\},
\end{equation}
where:
\begin{align}
{\bf{F}}_4={\left[1~1~1~0~1~1~1~0~0~0~0~1\right]}^{\mathrm{T}}\\
{\bf{S}}_4={\left[1~1~1~0~1~1~1~2~0~0~0~1\right]}^{\mathrm{T}}
\end{align}
\item The shared basic vector among $\mathrm{TX}_1$, $\mathrm{TX}_2$ and $\mathrm{TX}_4$ can be represented as follows ($\mathcal{Q}=\{1,2,4\}$):
\begin{equation}
\Big{\vert} {\bigcap_{q \in \mathcal{Q}}{\mathcal{V}^{[q]}}} \Big{\vert}=\Big{\lbrace} {{{\bf {F}}_{q'_1}} \Big{\vert} ~q'_l \in \{1,\dots,4\}-\mathcal{Q}} \Big{\rbrace}=\{{{\bf {F}}_{3}}\},
\end{equation}
where:
\begin{align}
{\bf{F}}_3={\left[1~1~0~1~1~1~0~1~0~0~1~0\right]}^{\mathrm{T}}\\
{\bf{S}}_3={\left[1~1~0~1~1~1~2~1~0~0~1~0\right]}^{\mathrm{T}}
\end{align}
\item The shared basic vector among $\mathrm{TX}_1$, $\mathrm{TX}_3$ and $\mathrm{TX}_4$ can be represented as follows ($\mathcal{Q}=\{1,3,4\}$):
\begin{equation}
\Big{\vert} {\bigcap_{q \in \mathcal{Q}}{\mathcal{V}^{[q]}}} \Big{\vert}=\Big{\lbrace} {{{\bf {F}}_{q'_1}} \Big{\vert} ~q'_l \in \{1,\dots,4\}-\mathcal{Q}} \Big{\rbrace}=\{{{\bf {F}}_{2}}\},
\end{equation}
where:
\begin{align}
{\bf{F}}_2={\left[1~0~1~1~1~0~1~1~0~1~0~0\right]}^{\mathrm{T}}\\
{\bf{S}}_2={\left[1~0~1~1~1~2~1~1~0~1~0~0\right]}^{\mathrm{T}}
\end{align}
\item The shared basic vector among $\mathrm{TX}_2$, $\mathrm{TX}_3$ and $\mathrm{TX}_4$ can be represented as follows ($\mathcal{Q}=\{2,3,4\}$):
\begin{equation}
\Big{\vert} {\bigcap_{q \in \mathcal{Q}}{\mathcal{V}^{[q]}}} \Big{\vert}=\Big{\lbrace} {{{\bf {F}}_{q'_1}} \Big{\vert} ~q'_l \in \{1,\dots,4\}-\mathcal{Q}} \Big{\rbrace}=\{{{\bf {F}}_{1}}\},
\end{equation}
where:
\begin{align}
{\bf{F}}_1={\left[0~1~1~1~0~1~1~1~1~0~0~0\right]}^{\mathrm{T}}\\
{\bf{S}}_1={\left[0~1~1~1~2~1~1~1~1~0~0~0\right]}^{\mathrm{T}}
\end{align}
\end{itemize}
From the above relations the basic vectors of different transmitted can be calculated as follows:
\begin{align}
{\bf{v}}^{[1]}_{1}={\bf{v}}^{[2]}_{1}={\bf{v}}^{[3]}_{1}={\bf F}_4={\left[1~1~1~0~1~1~1~0~0~0~0~1\right]}^{\mathrm{T}},\\
{\bf{v}}^{[1]}_{2}={\bf{v}}^{[2]}_{2}={\bf{v}}^{[4]}_{1}={\bf F}_3={\left[1~1~0~1~1~1~0~1~0~0~1~0\right]}^{\mathrm{T}},\\
{\bf{v}}^{[1]}_{3}={\bf{v}}^{[3]}_{2}={\bf{v}}^{[4]}_{2}={\bf F}_2={\left[1~0~1~1~1~0~1~1~0~1~0~0\right]}^{\mathrm{T}},\\
{\bf{v}}^{[2]}_{3}={\bf{v}}^{[3]}_{3}={\bf{v}}^{[4]}_{3}={\bf F}_1={\left[0~1~1~1~0~1~1~1~1~0~0~0\right]}^{\mathrm{T}}.
\end{align}
For the basic vectors of ${\bf{v}}^{[1]}_{1}={\bf{v}}^{[2]}_{1}={\bf{v}}^{[3]}_{1}$, since ${\bf{S}}_4={\left[1~1~1~0~1~1~1~2~0~0~0~1\right]}^{\mathrm{T}}$ the channel matrix ${\bar{\bf H}}^{[4q]}=\mathrm{diag}\left({\left[ h^{[4q]}(1)\dots h^{[4q]}(0) h^{[4q]}(1) \dots h^{[4q]}(2) h^{[4q]}(0) \dots h^{[4q]}(1)\right]}\right)$, which can not change the space spanned by the basic vectors ${\bf{v}}^{[1]}_{1}={\bf{v}}^{[2]}_{1}={\bf{v}}^{[3]}_{1}$ and these vectors remain align at $\mathrm{RX}_4$. Similar notion can be expressed for other basic vectors at different transmitters and receivers. 
\begin{lem}
{\it For the basic vectors of ${\bf v}^{[q]}_i=\bigcap _{q \in \mathcal{Q}}{\mathcal{V}^{[q]}}, \mathcal{Q}=\{q_1,\dots,q_r\}$, using switching pattern ${\bf S}_p$ at $\mathrm{RX}_p, p \in \mathcal{Q}$, the received basic vectors of $\bar{{\bf H}}^{[pq]}{\bf v}^{[q]}_i, q \in \mathcal{Q}$ at $\mathrm{RX}_p$ are linearly independent.}
\end{lem}
\begin{IEEEproof}
The basic vector ${\bf{v}}^{[q]}_{i}$ similar to \eqref{vectorrep} can be represented by the following equation:
\begin{equation}
{\bf{v}}^{[q]}_{i}={\left[{\left({\bf{v}}^{[q]}_{{{\bf e} i}}\right)^{\mathrm{T}}},\dots,{\left({\bf{v}}^{[q]}_{{{\bf e} i}}\right)^{\mathrm{T}}},{\left({\bf{v}}^{[q]}_{{{\bf f} i}}\right)^{\mathrm{T}}}\right]}^{\mathrm{T}}.
\end{equation}
If we show that at $\mathrm{RX}_p , p \in \mathcal{Q}$, all the $\bar{{\bf H}}^{[pq]}\left({1:(r-1)K}\right){\left[{\left({\bf{v}}^{[q]}_{{{\bf e} i}}\right)^{\mathrm{T}}},\dots,{\left({\bf{v}}^{[q]}_{{{\bf e} i}}\right)^{\mathrm{T}}}\right]}^{\mathrm{T}},q \in \mathcal{Q}$ are linearly independent, the proof will be accomplished. In this case all the nonzero elements of ${\left[{\left({\bf{v}}^{[q]}_{{{\bf e} i}}\right)^{\mathrm{T}}},\dots,{\left({\bf{v}}^{[q]}_{{{\bf e} i}}\right)^{\mathrm{T}}}\right]}^{\mathrm{T}}$ are in the sets $\{p_1,\dots,p_r\}$, $\{K+q_1,\dots,K+q_r\}$, ... and $\{(r-2)K+q_1,\dots,(r-2)K+q_r\}$. Also from \eqref{switchingpattern} all the first $(r-1)K$ elements of the channels which are connected to the $\mathrm{RX}_p, p \in \mathcal{Q}$ have the following form:
\begin{equation}
\begin{aligned}
&{\bar{\bf{H}}}^{[pq]}{\left(1:(r-1)K\right)}=&\\
&\mathrm{diag}\left({\left[{h^{[pq]}_1(1)},\dots,{h^{[pq]}_{q_1}(0)},\dots,{h^{[pq]}_{K+q_1}(2)},\dots,{h^{[pq]}_{q_1+(r-2)K}(r-1)},\dots,{h^{[pq]}_{(r-1)K}(1)}\right]}\right),&
\end{aligned}
\end{equation}
the common received basic vectors from $\mathrm{TX}_q, q\in \mathcal{Q}$ at $\mathrm{RX}_p, p \in \mathcal{Q}$ at least have $r$ different elements. Therefore, all the ${\bar{\bf{H}}}^{[pq]}{\left(1:(r-1)K\right)} {\bf v}^{[q]}_i \left(1:(r-1)K\right)$, $q,p \in \mathcal{Q}$ are linearly independent. So the proof is completed.
%\IEEEQEDhere
\end{IEEEproof}
 In the next section, we show that using the designed switching antenna pattern and the designed precoders, the $\frac{Kr}{r^2-r+K}$ sum DoF can be achieved.

\subsection{DoF achievability using the proposed switching pattern and the designed precoders}
Now we want to show that by the designed precoders the sum DoF of $\frac{Kr}{r^2-r+K}, r=\lceil{\frac{\sqrt{1+4K}-1}{2}}\rceil$ is achievable. In our designed precoders every transmitter e.g. $\mathrm{TX}_j$ has $\binom{K-1}{r-1}$ basic vectors. From Lemma 1 every generated basic vector at $\mathrm{TX}_j$ are linearly independent and the total number of dimensions used at each transmitter is equal to $\binom{K-1}{r-1}$. The received basic vectors at each receiver have two different types as follows:
\begin{enumerate}
\item The basic vectors which are linearly independent of each other.
\item The basic vectors which are aligned with each other.
\end{enumerate}
All the transmitted basic vectors which are linearly independent, because of $\lvert{h^{[pq]}_i}\rvert > 0,~p,q \in\{1,\dots,K\}$ also remain linearly independent at all the receivers. From the point of view of the $\mathrm{RX}_j$ and Lemma 3, the basic vectors which are not shared with the basic vectors of the $\mathrm{TX}_j$ are aligned with each other at the $\mathrm{RX}_j$. The number of such basic vectors can be calculated by counting $r$ different choosable transmitters among $K-1$ transmitters (except $\mathrm{TX}_j$) which is equal to $\binom{K-1}{r}$. Also, there are some basic vectors which are shared among $j^{th}$ transmitter and all other transmitters. The number of such vectors can be calculated by counting the number of $r-1$ choosable transmitters among $K-1$ ones which is equal to $\binom{K-1}{r-1}$. From Lemma 4 such basic vectors are linearly independent and therefore occupy $r\binom{K-1}{r-1}$ dimensions at $j^{th}$ receiver. Therefore, at $\mathrm{RX}_j$ the $\mathrm{TX}_j$ occupies $\binom{K-1}{r-1}$ dimensions (desired signal space dimensions) at its corresponding receiver. Also, we have $\left({r-1}\right)\binom{K-1}{r-1}$ dimensions which are generated by the basic vectors shared among $\mathrm{TX}_j$ and all other transmitters. These basic vectors from Lemma 4 are linearly independent and the total number of dimensions occupied by such vectors is $(r-1)\binom{K-1}{r-1}+\binom{K-1}{r-1}=r\binom{K-1}{r-1}$. Therefore the total number of dimensions is equal to summing $r\binom{K-1}{r-1}$ and $\binom{K-1}{r}$ dimensions which is equal to $r\binom{K-1}{r-1}+\binom{K-1}{r}$. The number of desired signal dimensions at $\mathrm{RX}_j$ is equal to $\binom{K-1}{r-1}$, which means that the total number of desired signal dimensions at $j^{th}$ user equals to $\binom{K-1}{r-1}$ from $r\binom{K-1}{r-1}+\binom{K-1}{r}$ total dimensions or transmission time slots. Consequently the DoF of $\frac{\binom{K-1}{r-1}}{r\binom{K-1}{r-1}+\binom{K-1}{r}}=\frac{r}{r^2-r+K}$ for $j^{th}$ user can be achievable. By the similar method of proof, we can show that all other transmitters can get to $\frac{r}{r^2-r+K}$ DoF and the $K-$user interference network totally can reach the sum DoF of $\frac{Kr}{r^2-r+K}$, which meets the upper-bound. Figure 3 shows DoF rate region of $K-$user interference channel using reconfigurable antenna. The result shows that the proposed method in \cite{Alaa} traces our method for $2 \leq K \leq 6$ and satisfies the sum DoF proposed by Wang in \cite{7}.   
\begin{figure}
  \centering
  \includegraphics[width=1\textwidth]%
    {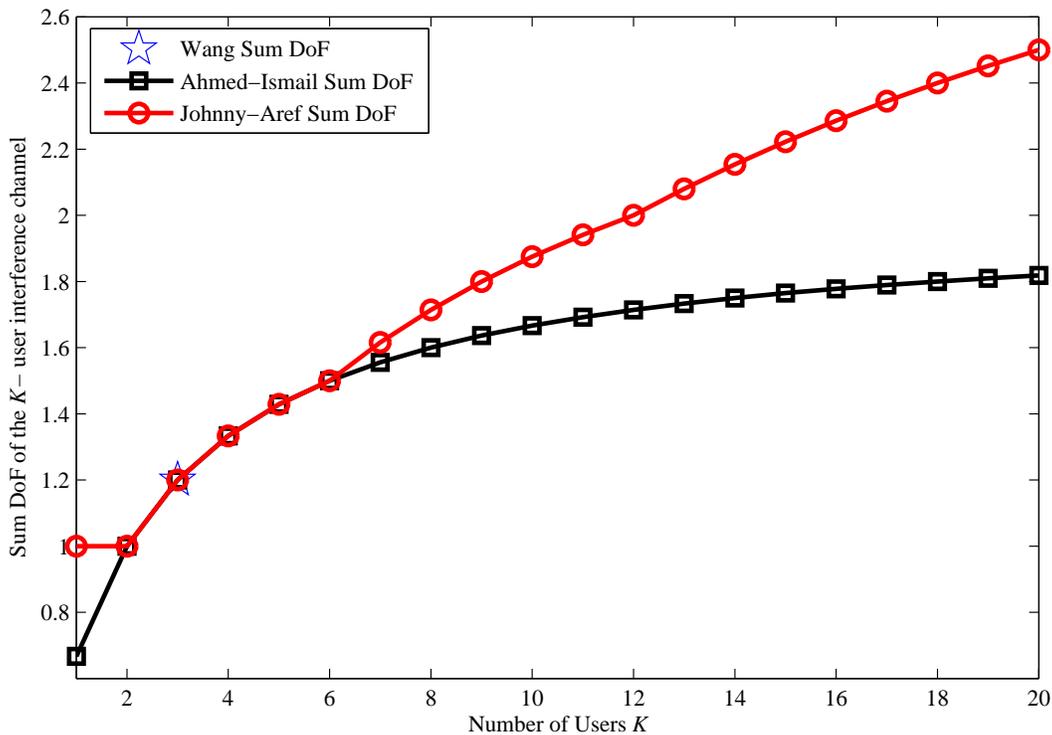}% picture filename
  \caption{Sum DoF of the $K-$user interference channel versus different number of the users $K$.}
\end{figure}      
\subsection{5-user SISO IC BIA, using reconfigurable antenna}
Consider a fully connected 5-user SISO Interference Channel. The maximum achievable sum DoF in this case can be found by setting $r=\left \lceil{\frac{\sqrt{1+4K}-1}{2}} \right \rceil{\vert_{K=5}}=2$ in the relation $\frac{Kr}{r^2-r+K}$ which is equal to $\frac{20}{14}> 1$. In this setting every transmitter can send 4 symbols through 14 time slots. In order to design precoders first of all we demonstrate the matrix $\bf F$ as follows:
\begin{equation}
\label{15}
{\bf {S}}^{\mathrm{T}}=
\left[ \begin{array}{cccccccccccccc}
         0 & 1 & 1 & 1 & 1 & 0 & 0 & 0 & 0 & 1 & 1 & 1 & 1 & 1\\
         1 & 0 & 1 & 1 & 1 & 0 & 1 & 1 & 1 & 0 & 0 & 0 & 1 & 1\\
         1 & 1 & 0 & 1 & 1 & 1 & 0 & 1 & 1 & 0 & 1 & 1 & 0 & 0\\
         1 & 1 & 1 & 0 & 1 & 1 & 1 & 0 & 1 & 1 & 0 & 1 & 0 & 1\\
         1 & 1 & 1 & 1 & 0 & 1 & 1 & 1 & 0 & 1 & 1 & 0 & 1 & 0
\end{array} \right]
\end{equation}    
In this case since $r=2$, the value of the matrix $\bf S=\bf F$. Also, from \eqref{preco}, we can design all the $\binom{5}{2}=10$ basic vectors at each transmitter as follows:
\begin{align}
&{\bf v}_1^{[1]}={\bf v}_1^{[2]}={\left[ {1~1~0~0~0~1~0~0~0~0~0~0~0~0}\right]}^{\mathrm{T}}&\\
&{\bf v}_2^{[1]}={\bf v}_1^{[3]}={\left[ {1~0~1~0~0~0~1~0~0~0~0~0~0~0}\right]}^{\mathrm{T}}&\\
&{\bf v}_3^{[1]}={\bf v}_1^{[4]}={\left[ {1~0~0~1~0~0~0~1~0~0~0~0~0~0}\right]}^{\mathrm{T}}&\\
&{\bf v}_4^{[1]}={\bf v}_1^{[5]}={\left[ {1~0~0~0~1~0~0~0~1~0~0~0~0~0}\right]}^{\mathrm{T}}&\\
&{\bf v}_2^{[2]}={\bf v}_2^{[3]}={\left[ {0~1~1~0~0~0~0~0~0~1~0~0~0~0}\right]}^{\mathrm{T}}&\\
&{\bf v}_3^{[2]}={\bf v}_2^{[4]}={\left[ {0~1~0~1~0~0~0~0~0~0~1~0~0~0}\right]}^{\mathrm{T}}&\\
&{\bf v}_4^{[2]}={\bf v}_2^{[5]}={\left[ {0~1~0~0~1~0~0~0~0~0~0~1~0~0}\right]}^{\mathrm{T}}&\\
&{\bf v}_3^{[3]}={\bf v}_3^{[4]}={\left[ {0~0~1~1~0~0~0~0~0~0~0~0~1~0}\right]}^{\mathrm{T}}&\\
&{\bf v}_4^{[3]}={\bf v}_3^{[5]}={\left[ {0~0~1~0~1~0~0~0~0~0~0~0~0~1}\right]}^{\mathrm{T}}&\\
&{\bf v}_4^{[4]}={\bf v}_4^{[5]}={\left[ {0~0~0~1~1~0~0~0~0~0~0~0~0~0}\right]}^{\mathrm{T}}.&
\end{align} 
As it was proved in Lemma 2, all the generated basic vectors at each transmitter are linearly independent e.g. ${\bf v}_1^{[1]}$, ${\bf v}_2^{[1]}$, ${\bf v}_3^{[1]}$ and ${\bf v}_4^{[1]}$ at $\mathrm{TX}_1$. Now we can design the switching pattern at each receiver. In this case since the optimum value of $r$ is equal to 2, every receiver is equipped with an antenna with two RF chains or switching modes. Therefore, each receiver during data reception can switch between its two RF chains. From \eqref{switchingpattern} we can get switching pattern at each receiver as follows:
\begin{align}
&{\bf S}_1=\left[0~1~1~1~1~0~0~0~0~1~1~1~1~1\right]^{\mathrm{T}}&\\
&{\bf S}_2=\left[1~0~1~1~1~0~1~1~1~0~0~0~1~1\right]^{\mathrm{T}}&\\
&{\bf S}_3=\left[1~1~0~1~1~1~0~1~1~0~1~1~0~0\right]^{\mathrm{T}}&\\
&{\bf S}_4=\left[1~1~1~0~1~1~1~0~1~1~0~1~0~1\right]^{\mathrm{T}}&\\
&{\bf S}_5=\left[1~1~1~1~0~1~1~1~0~1~1~0~1~0\right]^{\mathrm{T}}.&
\end{align} 
In this case due to the above switching pattern, for $\mathrm{RX}_1$, we have the following channel realization:
\begin{equation}
{\bar {\bf H}}^{[1q]}=\mathrm{diag}\left({\left[{h_1^{[1q]}(0),h_2^{[1q]}(1),\dots,h_5^{[1q]}(1),h_6^{[1q]}(0),\dots,h_9^{[1q]}(0),h_{10}^{[1q]}(1),\dots,h_{14}^{[1q]}(1)}\right]}\right).
\end{equation}
Therefore, the members of the set $\mathcal{S}^{[1]}$, shows the basic vectors which span the space of the first receiver:
\begin{equation}
\begin{aligned}
\mathcal{S}^{[1]}=& \Big{\lbrace}\underbrace{{\bar{\bf H}}^{[11]}{\bf v}_1^{[1]},{\bar{\bf H}}^{[12]}{\bf v}_1^{[2]}}_{\text{linearly independent}},\underbrace{{\bar{\bf H}}^{[11]}{\bf v}_2^{[1]},{\bar{\bf H}}^{[13]}{\bf v}_1^{[3]}}_{\text{linearly independent}},\underbrace{{\bar{\bf H}}^{[11]}{\bf v}_3^{[1]},{\bar{\bf H}}^{[14]}{\bf v}_1^{[4]}}_{\text{linearly independent}},\underbrace{{\bar{\bf H}}^{[11]}{\bf v}_4^{[1]},{\bar{\bf H}}^{[12]}{\bf v}_1^{[5]}}_{\text{linearly independent}}&\\
&\underbrace{{\bar{\bf H}}^{[12]}{\bf v}_2^{[2]},{\bar{\bf H}}^{[13]}{\bf v}_2^{[3]}}_{\text{align}},\underbrace{{\bar{\bf H}}^{[12]}{\bf v}_3^{[2]},{\bar{\bf H}}^{[13]}{\bf v}_2^{[4]}}_{\text{align}},\underbrace{{\bar{\bf H}}^{[11]}{\bf v}_4^{[2]},{\bar{\bf H}}^{[15]}{\bf v}_2^{[5]}}_{\text{align}},\underbrace{{\bar{\bf H}}^{[12]}{\bf v}_3^{[3]},{\bar{\bf H}}^{[12]}{\bf v}_3^{[4]}}_{\text{align}}&\\
&\underbrace{{\bar{\bf H}}^{[12]}{\bf v}_4^{[3]},{\bar{\bf H}}^{[13]}{\bf v}_3^{[5]}}_{\text{align}},\underbrace{{\bar{\bf H}}^{[12]}{\bf v}_4^{[4]},{\bar{\bf H}}^{[13]}{\bf v}_4^{[5]}}_{\text{align}} \Big{\rbrace}.&
\end{aligned}
\end{equation} 
Since $\bar{\bf H}^{[1q]}$, in the time slots of $\{2,3,4,5,10,11,12,14\}$ and $\{1,6,7,8,9\}$ experiences similar coefficients of $h^{[1q]}(1)$ and $h^{[1q]}(0)$ respectively, the basic vectors of ${\bf v}^{[j]}_i, i>1, j \neq 1$ are aligned with ${\bar{\bf H}}^{[1j]}{\bf v}^{[j]}_i, i>1, j \neq 1$. In other words, in this case we have:
\begin{equation}
\mathrm{dim}\left({\left[ {{\bar{\bf H}}^{[1j]}{\bf v}^{[j]}_i~{\bf v}^{[j]}_i} \right] }\right)=1,~i>1, j \neq 1.
\end{equation}      
The above relation shows that all the shared generated basic vectors such as $\lbrace{\bf v}_2^{[2]},{\bf v}_2^{[3]}\rbrace$, $\{{\bf v}_3^{[2]},{\bf v}_2^{[4]}\}$, $\{{\bf v}_4^{[2]},{\bf v}_2^{[5]}\}$, $\{{\bf v}_3^{[3]},{\bf v}_3^{[4]}\}$, $\{{\bf v}_4^{[3]},{\bf v}_3^{[5]}\}$ and $\{{\bf v}_4^{[4]},{\bf v}_4^{[5]}\}$ after being multiplied by channel matrices of ${\bar{\bf H}}^{[1j]}, j \neq 1$ remain aligned with each other. In this case since the basic vectors of $\{{\bf v}_1^{[1]},{\bf v}_1^{[2]}\}$ have the nonzero elements in the time slots of $\{1,2,6\}$ and the channel model matrix changes its value between time slots of one and two, both ${\bar{\bf H}}^{[11]}{\bf v}_1^{[1]}$ and ${\bar{\bf H}}^{[12]}{\bf v}_1^{[2]}$ are linearly independent. Similarly all other received basic vectors of $\{\bar{\bf H}^{[11]}{\bf v}_2^{[1]},\bar{\bf H}^{[13]}{\bf v}_1^{[3]}\}$, $\{\bar{\bf H}^{[11]}{\bf v}_3^{[1]},\bar{\bf H}^{[14]}{\bf v}_1^{[4]}\}$ and $\{\bar{\bf H}^{[11]}{\bf v}_4^{[1]},\bar{\bf H}^{[15]}{\bf v}_1^{[5]}\}$ are jointly linearly independent. Therefore, at the first receiver from 14 dimensions we have four free interference dimensions and this user can achieve $\frac{4}{14}$ DoF. Similarly we can achieve $\frac{4}{14}$ for all other users and totally we get $\frac{10}{7}$ sum DoF. 

\section{Conclusion}
In this paper, we have shown that in the $K$-user SISO interference channel the sum DoF of the linear BIA using reconfigurable antenna is $\max_{r \in \mathbb{N}}\frac{Kr}{r^2-r+K}$. We provide both achievability and converse proof for this important problem. A key insight is that each signal dimension from one user can be aligned with a set of distinct transmitters at the receivers with complimentary set.  Without channel state information at the transmitters, this result indicates that when the value of $K$ limits to infinity we can achieve $\frac{\sqrt{K}}{2}$ compared to the unity achievable DoF of the orthogonal multiple access schemes. Moreover, in achievability sections we proposed an algorithm to generate the transmit beamforming vectors and antenna switching patterns utilized in BIA. We showed that the proposed algorithm can achieve the $\frac{Kr}{r^2-r+K}$ sum DoF for any $K$ and $r = \left \lceil{\frac{\sqrt{1+4K}-1}{2}} \right \rceil$ values. Also we show that the term $\frac{Kr}{r^2-r+K}$ is maximized when the value of $r \in \mathbb{N}$ is equal to $\left \lceil{\frac{\sqrt{1+4K}-1}{2}} \right \rceil$. By applying both achievability method and converse proof of this work for the 3-user Interference Channel, we showed that a sum DoF of $\frac{6}{5}$, which was obtained previously in \cite{7} was met. Using designed switching pattern assumptions has important hardware implications. For instance, the proposed algorithm operates with low cost reconfigurable antennas that have only $r$ modes and there is no need for transmitters to have access to channel CSI. Also the structure of beamforming vectors is very simple and can be applicable by activating or deactivating certain symbols at the transmitters. 
\appendices
%\section{Proof of the inequality}
%\begin{lem}{\it For all values of $K$ and $r=\lceil {\frac{\sqrt{1+4K}-1}{2}} \rceil$, we can conclude that $\left(n-(r-1)K\right)-\binom{K-1}{r-1} \geq 0$.}
%\end{lem}
%\begin{IEEEproof}
%Starting from finding the sign of the term $\left(n-(r-1)K\right)-\binom{K-1}{r-1}$, we have:
%\begin{align}
%\mathrm{sgn} &\left(r \binom{K-1}{r-1}+\binom{K-1}{r}-(r-1)K-\binom{K-1}{r-1}\right )&\\
%&=\mathrm{sgn}\left({(r-1)\binom{K-1}{r-1}+\binom{K-1}{r}}-(r-1)K\right)&
%\end{align}
%For $3\leq K \leq 6$ where, $r=2$ the term ${{(r-1)\binom{K-1}{r-1}+\binom{K-1}{r}}-(r-1)K}$ can be simplified as follows:
%\begin{equation}
%{\binom{K-1}{2}-1}\geq 0,3\leq K \leq 6,
%\end{equation}
%which satisfies the result of this lemma. For $K > 6$, by the use of Stirling's approximation, we can easily show that the term $\binom{K-1}{r-1}$ is strictly larger than the value of $K$. Therefore, the term ${\binom{K-1}{r-1}-K}$ is surely larger than zero and the proof of this lemma is completed.
%\end{IEEEproof}
\ifCLASSOPTIONcaptionsoff
  \newpage
\fi
%\IEEEtriggeratref{6}

\end{document}